\documentclass[
								reprint,
								aps,
								prl,
								nofootinbib,
								twocolumn,
								notitlepage,
							]{revtex4-1}

\usepackage{lmodern}
\usepackage{amsmath,amssymb,amsthm,amsfonts,mathrsfs}
\usepackage{graphicx,epstopdf,epsfig,enumerate}
\usepackage{color}
\usepackage[colorlinks=true,linkcolor=blue,citecolor=blue,urlcolor=blue]{hyperref}

\DeclareGraphicsExtensions{.eps}

\renewcommand{\vec}[1]{\mathbf{#1}}

\begin{document}

\title{Thermoelectric generators as  self-oscillating heat engines}

\author{Robert Alicki}
\email[e-mail: ]{fizra@ug.edu.pl}
\affiliation{Institute of Theoretical Physics and Astrophysics, University of Gda\'nsk, 80-952 Gda\'nsk, Poland}

\begin{abstract}
In the previous paper \cite{Alicki:2015} a model of a solar cell has been proposed in which the non-periodic source of energy - photon flux - drives the collective periodic motion of electrons in a form of plasma oscillations. Subsequently, plasma oscillations are rectified by the p-n junction diode into the direct current (work). This approach makes a solar cell similar to standard macroscopic heat motors or turbines  which always contain two heat baths, the working medium and the periodically moving piston or rotor. Here, a very similar model is proposed in order to describe the operation principles of  thermoelectric generators  based either on bimetallic or semiconductor p-n junctions. Again plasma oscillation corresponds to a piston and sunlight is replaced by a hot bath. The mathematical formalism is based on the  Markovian master equations which can be derived in a rigorous way from the underlying Hamiltonian models and are consistent with the laws of thermodynamics.
\end{abstract}

\maketitle

\section{Introduction}

While the macroscopic heat motors or turbines contain always  periodically moving elements like pistons, flywheels or rotors \cite{Jenkins},\cite{AVK}, the photovoltaic devices, thermoelectric generators or biological  engines driven either by sunlight (photosynthesis) or chemical energy seem, according to the standard wisdom, to avoid any self-oscillation mechanisms. On the other hand extensive theoretical studies of quantum heat engines within the formalism of quantum open systems \cite{Alicki:1979} -\cite{Kosloff} show that the process of work extraction must involve a work reservoir being a system of a single degree of freedom and executing an oscillatory motion which in the semi-classical limit can be replaced by the periodic external driving. The presence of such a system is even necessary to properly define the notion of work as a deterministic form of energy. It suggests that for the examples mentioned above a certain ``hidden" self-oscillation mechanism of work generation must also be present. Indeed, as already noticed the standard picture of current generation as caused by the emerging electric field in a junction cannot be correct because a DC current cannot be driven in a closed circuit by a purely electrical potential difference \cite{Wuerfel}. In the previous paper \cite{Alicki:2015} it was shown that the consistent description of work generation by a solar cell can be formulated introducing a work reservoir in a form of THz plasma oscillations. 
\par
The aim of the present contribution is to show that a very similar model can explain the operation principles of thermoelectric generators. The basic difference between semiconductor and bimetallic devices concerns the frequency of plasma oscillations. For  doped semiconductors plasma frequencies are in the THz domain (low infrared) \cite{plasma1}, \cite{plasma2} and  can be treated as slow while for metals they correspond to the ultraviolet region (PHz = $10^{15}$ Hz) and therefore are too fast to be effectively coupled to the thermal motion of a lattice. However, it is argued in the Appendix, using hydrodynamical model of electron gas, that under reasonable assumptions  there should exist  THz modes of plasma oscillations in bimetallic contacts.
\section{Heat engine with a slow piston }
Consider a simplified version of the generic quantum  heat engine model introduced in \cite{Alicki:1979} and successfully used in \cite{Alicki:2015} in the context of solar cells. The system, corresponding to the ``working medium " interacts weakly with  two heat baths at different temperatures. The macroscopic piston is described by the external driving  $V(t)$ added to the free Hamiltonian $H_0$ of the system. 
\par
For  slowly varying $V(t)$ (in comparison to the fast internal dynamics )   the irreversible evolution of the system's time-dependent reduced density matrix $\rho(t)$ satisfies the following Markovian master equation  (MME) 
\begin{equation}
\frac{d}{dt}\rho(t)  = -i[H(t),\rho(t)] + \mathcal{L}(t)\rho(t) ,
\label{master2}
\end{equation}
where $H(t) = H_0 + V(t)$ is the total Hamiltonian of the system and $\mathcal{L}(t)$ describes the action of the baths on the system ( $\hbar = k_B =1$ ) .
\par

\par
The periodic driving executed by a large semi-classical oscillator commutes with the system Hamiltonian and reads
\begin{equation}
V(t) = g(\sin\Omega t)\, M, \quad [H_0 , M] = 0,
\label{driving}
\end{equation}
where $g \ll 1$ is a small amplitude of oscillations.
The dissipative generator $\mathcal{L}(t)$ obtained by the weak coupling limit procedure is a function of the magnitude of perturbation and can be written as 
\begin{equation}
\mathcal{L}(t) \equiv \mathcal{L}[\xi(t)], \quad \xi(t) = g\sin\Omega t .
\label{generator}
\end{equation}
For all $\xi$ the generators $\mathcal{L}[\xi]$ commute and possess the stationary states $\bar{\rho}[\xi]$ satisfying the identities
\begin{equation}
\mathcal{L}[\xi]\bar{\rho}[\xi] = 0,  \quad \mathcal{L}'[\xi]\bar{\rho}[\xi]  =-\mathcal{L}[\xi]\bar{\rho}'[\xi]  
\label{identity}
\end{equation}
where $\mathcal{L}'[\xi] \equiv \frac{d}{d\xi}\mathcal{L}[\xi]$ , $\bar{\rho}'[\xi] \equiv \frac{d}{d\xi}\bar{\rho}[\xi] $.
\subsection{Formula for power}
The power $P(t)$ provided by the engine and the net heat current $J(t)$ supplied by the baths are defined as \cite{Alicki:1979}
\begin{equation}
P(t) = -\mathrm{Tr}\Bigl(\rho(t) \frac{d H(t)}{dt}\Bigr),\quad 
J(t) = \mathrm{Tr}\Bigl(H(t) \frac{d \rho(t)}{dt}\Bigr).
\label{work_heat}
\end{equation}
Those definitions are the only ones which are consistent with the first and second law of thermodynamics  and
the time-dependence of the Hamiltonian is necessary to define work. The stationary average power output  of the engine reads
\begin{equation}
\bar{P} = -g\Omega\lim_{t_0\to\infty} \frac{1}{t_0}\int_0^{t_0}\mathrm{Tr}\Bigl(\rho(t) M \Bigr)\cos\Omega t\,dt .
\label{power}
\end{equation}
Expanding the average power  with respect to $g$ one obtains the second order approximation \cite{Alicki:2015}
\begin{equation}
\bar{P} = -\frac{1}{2}g^2 \mathrm{Tr}\Bigl(\bar{\rho}'[0] \frac{\Omega^2}{\Omega^2 + {\mathcal{L}^*}^2[0]}{\mathcal{L}^*}[0]M\Bigr) ,
\label{power1}
\end{equation}
where ${\mathcal{L}^*}[0]$ is the Heisenberg picture counterpart of the Schroedinger picture generator $\mathcal{L}[0]$. To derive the final expression one assumes that, although the modulation is slow with respect to the intrinsic motion of the system, its frequency is much higher than the relaxation rate of the observable $M$ hidden in  ${\mathcal{L}^*}[0]$. Then, we have
\begin{equation}\label{power2}
\bar{P} = -\frac{1}{2}g^2 \mathrm{Tr}\Bigl(\bar{\rho}'[0] \mathcal{L}^*[0]M\Bigr) .
\end{equation}
The lowest order formula \eqref{power2} is still consistent with thermodynamics \cite{Alicki:2015} and is the basic one for the further analysis of  thermoelectric generators.
Notice, that the stationary output power is proportional to the square of the amplitude of piston oscillations. This amplitude is a free parameter which is determined by the energy flux from the hot bath and the load attached to the oscillator (e.g. the resistance of the external electric circuit).
\section{Model of thermoelectric generator}
The model of thermoelectric generator is illustrated on Fig.1. The working medium is an electron gas distributed in two boxes $A$ and $B$, corresponding to different materials, metals or doped semiconductors, connected by a junction. The electrostatic  potential jump through the junction denoted by $E_g$ is typically of the order of few $ eV$. We assume that in the case of metals the density of electrons in the box $A$ is lower than in $B$ what implies that potential energy of the electron in the box $A$ is higher than in the box $B$. For semiconductors the box $A$ is a p-type and the box $B$ a n-type doped material. The hot bath interacts with the electrons in the interface region influencing their transport through the junction while the cold bath at the ambient temperature cools down the electrons in the bulk. In the junction region an interface between two different material is formed with the local concentration of charges and build-in electrostatic  potential. One assumes the existence of collective charge oscillations localized at the junction with typical frequencies in THz domain. For semiconductor devices such oscillations are experimentally confirmed and interpreted as plasma oscillations with the plasma frequency 
\begin{equation}
\omega_p = \sqrt{\frac{ne^2}{m_* \epsilon_0}}, 
\label{plasma_f}
\end{equation}
where $n$ is a density of charge carriers and $m_*$ their effective mass. The bimetallic junction is more complicated as the bulk and surface  plasma oscillations in metals possess frequencies in the PHz domain ($10^{15} Hz$) and hence  cannot be effectively excited by thermal phonons. In the Appendix  the mechanism of slow  coherent charge oscillations is proposed which is related to the particular physical structure of  metal-metal junctions.

\begin{figure}[tb]
	\centering
\includegraphics[width=\columnwidth,keepaspectratio]{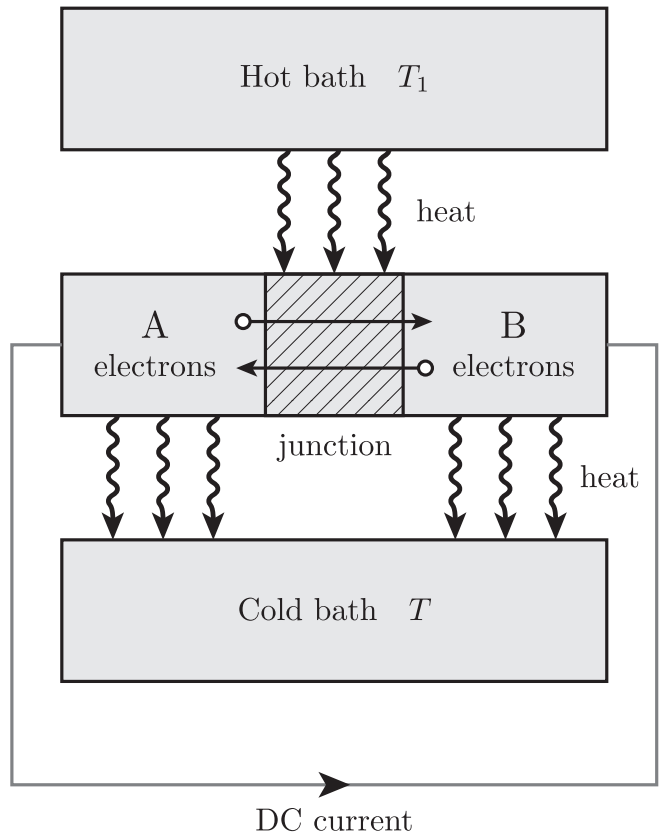}
	\caption{\textbf{Model of thermoelectric generator.} Bimetallic or p-n junction is heated by a hot bath increasing the transition rate between both regions. The electrons in bulk are cooled down by a cold bath. Rectified plasma oscillations produce a net flow of electrons from the region with their higher density (box B) to the lower one (box A).}
	\label{fig:Figure1}
\end{figure}
The electrons in two boxes $A$ and $B$ are described by two sets of the annihilation and  creation operators $a_{\vec{k}}$, $a^{\dagger}_{\vec{k}}$ and $b_{\vec{l}}$, $b^{\dagger}_{\vec{l}}$, respectively, subject to canonical anticommutation relations.
The electrons are treated as  non-interacting fermions moving in a self-consistent potential with the unperturbed Hamiltonian  
\begin{equation}
H_{0} =  \sum_{\vec{k}} E_{a}(\vec{k}) a^{\dagger}_{\vec{k}} a_{\vec{k}} +  \sum_{\vec{l}} E_{b}(\vec{l}) b^{\dagger}_{\vec{l}} b_{\vec{l}}. 
\label{ham_electrons}
\end{equation}
In a junction a non-homogeneous free carrier distribution created in a self-consistent build-in potential  can be perturbed producing collective plasma oscillations with the frequency $\Omega$. Those oscillations modulate periodically the Hamiltonian \eqref{ham_electrons} but the detailed mechanism is slightly different for the semiconductor and bimetallic junctions (see Appendix for the derivations). In both cases the associated time-dependent perturbation  which should be added to the electronic Hamiltonian \eqref{ham_electrons},  is proportional to the difference of electron densities in the boxes $B$ and $A$ and has a mean-field form ($\xi(t) = g\sin\Omega t$)

\begin{equation} 
\xi M = {\xi} E_g V_J \Bigl(\frac{1}{V_B}\sum_{\vec{l}}  b^{\dagger}_{\vec{l}} b_{\vec{l}} - \frac{1}{V_A}\sum_{\vec{k}}  a^{\dagger}_{\vec{k}} a_{\vec{k}}\Bigr) .
\label{ham_pert1}
\end{equation}
Here, $\xi$ is a small dimensionless parameter describing the magnitude of deformation, $V_A , V_B$ volumes of the boxes, $V_J$ is the effective volume of the junction region and $E_g$ is the relevant energy scale chosen here to be the potential jump across the junction. Remember, that for the semiconductor p-n junction the box $A$ corresponds to the ``p" part and $B$ to the ``n" part. In this case $-a^{\dagger}_{\vec{k}} a_{\vec{k}}$ should be rather interpreted as the number operator of holes minus an irrelevant constant.
\par
The interaction with two independent heat bath is described by the sum of two terms: $\mathcal{L}_{c}$ for the cold bath at the ambient temperature $T$ and $\mathcal{L}_{h}$ for the hot bath at the temperature $T_1 > T$. The coupling to the cold bath thermalizes electrons in both boxes independently and hence preserves separately both electron numbers
$ \sum_{\vec{k}}  a^{\dagger}_{\vec{k}} a_{\vec{k}}$ and $ \sum_{\vec{k}}  b^{\dagger}_{\vec{k}} b_{\vec{k}}$ . It means that
 $\mathcal{L}_{c}^*M = 0$ and $\mathcal{L}_{c}^*$ disappears from the formula \eqref{power2} for the stationary power output. Notice, that in the discussed idealized  model heat transport leading to a temperature gradient is neglected what corresponds to the infinite heat conductivity of the bulk.

\subsection{Hot bath generator and stationary state}
The generator $\mathcal{L}_{h}$  describes  thermally induced electron transitions from one box to another across the potential barrier and can be written in the following form  
\begin{equation} 
\mathcal{L}_{h}[0]  = \sum_{\vec{k}\vec{l}}\mathcal{L}^{(h)}_{\vec{k}\vec{l}}
\label{ME_hot}
\end{equation}
where
\begin{align}
		&\mathcal{L}^{(h)}_{\vec{k}\vec{l}}\rho = \gamma_{\vec{k}\vec{l}}\bigl( a_{\vec{k}} b_{\vec{l}}^{\dagger} \, \rho \,b_{\vec{l}} a_{\vec{k}}^{\dagger}  - \frac{1}{2} \{ b_{\vec{l}}a_{\vec{k}}^{\dagger} a_{\vec{k}} b_{\vec{l}}^{\dagger} ,\, \rho \}\bigr) \nonumber \\
		&+ e^{-E_g/T_1}\gamma_{\vec{k}\vec{l}} \bigl( a_{\vec{k}}^{\dagger} b_{\vec{l}} \, \rho \,  b_{\vec{l}}^{\dagger} a_{\vec{k}} - \frac{1}{2} \{ b_{\vec{l}}^{\dagger} a_{\vec{k}} a_{\vec{k}}^{\dagger} b_{\vec{l}} ,\, \rho \}\bigr), 
		\label{ME_hot1}
	\end{align}
The single term \eqref{ME_hot1} describes the following processes:\\

a) electron transfer from the box $A$ to $B$ accompanied by a  positive energy release to the hot bath equal to 
\begin{equation} 
E_{a}(\vec{k}) - E_{b}(\vec{l}) \simeq E_g,
\label{energy}
\end{equation}

b) the inverse process of electron transfer form the lower energy state in $B$ to the higher energy one in $A$ with the probability suppressed by the Boltzmann factor.
\par
The condition \eqref{energy} follows from the approximate energy conservation valid for the tunneling process through a barrier of the height $E_g$ assisted by  thermal fluctuations of the hot bath. It means also that the relaxation rates $\gamma_{\vec{k}\vec{l}}$ are essentially different from zero only if \eqref{energy} is satisfied. They are proportional to the spectral density of the hot bath at $\omega_{\vec{k}\vec{l}} = E_{a}(\vec{k}) - E_{b}(\vec{l}) \simeq E_g$.
\par
Although the generator $\mathcal{L}_{c}^*$ is absent in the formula for power the thermal relaxation of electrons in a given box is the fastest process and hence determines the form of the stationary state $\bar{\rho}$. Within a reasonable approximation one can assume that the stationary state of the electronic systems with the total Hamiltonian $H_0 + \xi M$  is a product of grand canonical ensembles for electrons in both boxes with the same temperature $T$ of the device
and different electro-chemical potentials $\mu_a$ and $\mu_b$, respectively. The associated density matrix has form
\begin{align}
\bar{\rho}[\xi] = &\frac{1}{Z[\xi]} \exp\left\{ -\frac{1}{T} \sum_{\vec{k}} \left[ \left(E_a(\vec{k})+ \xi E_g\frac{V_J}{V_A} -\mu_a\right) a^{\dagger}_{\vec{k}}a_{\vec{k}} \right. \right. \nonumber \\
&\left. \left. + \left(E_b(\vec{k}) - \xi E_g\frac{V_J}{V_B}-\mu_b\right) b^{\dagger}_{\vec{k}}b_{\vec{k}}\right]\right\} . 
\label{grand}
\end{align}
The electro-chemical potentials $\mu_a , \mu_b$ are determined by the numbers of carriers and hence by densities of electrons in both boxes. The difference of electro-chemical potentials is related to the measured voltage $\Phi$ between $A$ and $B$ 
\begin{equation} 
\mu_a - \mu_b = e \Phi .
\label{voltage}
\end{equation}
\subsection{Power and efficiency}
One can insert all elements computed in the previous section into the expression for power \eqref{power2}. Then one uses the properties of the quasi-free (fermionic Gaussian) stationary state \eqref{grand} which allow to reduce the averages of even products of annihilation and creation fermionic operators into sums of products of the only non-vanishing two-point correlations
\begin{subequations}
\begin{align}
	&\langle a_{\vec{k}}^{\dagger}a_{\vec{k}^{\prime}} \rangle_0 = \delta_{\vec{k}\vec{k}^{\prime}} f_{c}(\vec{k}),\quad 
	\langle a_{\vec{k}}a_{\vec{k}^{\prime}}^{\dagger}\rangle_0 = \delta_{\vec{k}\vec{k}^{\prime}}\left(1- f_{c}(\vec{k})\right), \nonumber \\
	&\langle b_{\vec{l}}^{\dagger}b_{\vec{l}^{\prime}}\rangle_0 = \delta_{\vec{l}\vec{l}^{\prime}} f_{b}(\vec{l}),\quad 
	\langle b_{\vec{l}}b_{\vec{l}^{\prime}}^{\dagger}\rangle_0 = \delta_{\vec{l}\vec{l}^{\prime}}\left(1- f_{b}(\vec{l})\right).\nonumber 
\end{align}
\end{subequations}
Here $\langle\cdots\rangle_0$ denotes the quantum average with respect to the state $\bar{\rho}[0]$ given by \eqref{grand}, and $f_{a}(\vec{k})$ and $f_{b}(\vec{l})$ are the Fermi-Dirac statistical distribution functions 
\begin{equation}
f_{a}(\vec{k}) = \frac{1}{e^{\beta (E_{a}(\vec{k})-\mu_{a})}+1}, \quad
f_{b}(\vec{l}) = \frac{1}{e^{\beta (E_{b}(\vec{l})-\mu_{b})}+1},
\end{equation}
with $\beta = 1/T$. 
Inserting the elements defined above into the general formula \eqref{power2} one obtains the leading order contribution to power in a following form
\begin{align}\label{power3a}
&\bar{P} = \frac{g^2 E_g^2 V_J^2 }{T} (V_A + V_B)(\bar{n}_b - \bar{n}_a )\, \times\\ 
&\frac{1}{V_A V_B}\sum_{\vec{k}\vec{l}}\gamma_{\vec{k}\vec{l}} \Bigl( e^{-E_g/T_1}\bigl[ 1 -f_{a}(\vec{k})\bigr] f_{b}(\vec{l}) 
- \bigl[ 1 -f_{b}(\vec{l})\bigr] f_{a}(\vec{k})\Bigr), \nonumber
\end{align}
where $\bar{n}_a = \frac{1}{V_A}\langle\sum_{\vec{k}} a_{\vec{k}}^{\dagger}a_{\vec{k}} \rangle_0 $, $\bar{n}_b = \frac{1}{V_B}\langle\sum_{\vec{l}} b_{\vec{l}}^{\dagger}b_{\vec{l}} \rangle_0 $ are electron densities. Using \eqref{voltage} one can rewrite \eqref{power3a} as
\begin{align}\label{power4}
\bar{P} = &\frac{g^2  E_g^2 V_J^2}{T}(V_A + V_B)(\bar{n}_b - \bar{n}_a )\Gamma\, \times \\
& \Bigl(\exp\Bigl\{\frac{1}{T}\Bigl(\Bigl[ 1-\frac{T}{T_1}\Bigr]E_g - e\Phi\Bigl)\Bigr\} -1\Bigr), \nonumber
\end{align}
where $\Gamma = \frac{1}{V_A V_B}\sum_{\vec{k}\vec{l}}\gamma_{\vec{k}\vec{l}}\bigl[ 1 -f_{b}(\vec{l})\bigr] f_{a}(\vec{k}) > 0$ is finite in the thermodynamical limit.
\par
The condition for work generation in the discussed model of idealized thermoelectric device  reads ($\Delta T = T_1 - T$)
\begin{equation}
e\Phi <  e\Phi_0 = E_g \Bigl( 1-\frac{T}{T_1}\Bigr) = \frac{E_g}{T_1} \Delta T 
\label{positive_work}
\end{equation}
The inequality in \eqref{positive_work} implies that $\Phi_0$ can be interpreted as an \emph{open-circuit voltage} of the device  and $\frac{E_g}{e T_1}$ as the \emph{relative Seebeck coefficient}. \\  
By closing  the external circuit one reduces the voltage and hence the output power is strictly positive driving collective charge oscillations (positive feedback). The charge oscillations are subsequently rectified by the diode mechanism of the junction producing a direct current.
\par
The presence of the Carnot factor $1 - T/T_1$ suggests also the interpretation of the eq. \eqref{positive_work} in terms of thermodynamical efficiency. Indeed, the transport of a single electron from the box $B$ to $A$ through the junction requires at least $E_g$ of thermal energy extracted from the hot bath.  Then, a part of energy $E_g $ is transformed into useful work, equal at most $e\Phi_0$ per single electron flowing in the external circuit.  
\par
For real systems the efficiency is much smaller than the Carnot bound because of the finite heat conductivity
and damping of plasma oscillations. Similarly, for a typical $T_1 \sim 500 K$, the relative Seebeck coefficient given by \eqref{positive_work} is of the order of $mV/K$ what is comparable to the highest values obtained for some semiconductors. For metals, the neglected irreversible transport processes reduce  the  Seebeck coefficient by two or three orders of magnitude.
\section{Conclusions} The aim of this paper is to show that the idea of ``hidden self-oscillations" proposed in \cite{Alicki:2015} for  solar cells can be  extended to thermoelectric generators. It is illustrated by the idealized model  in which heat conduction is neglected. The ``moving piston" is again represented by collective charge oscillations with frequencies in the THz domain. While such oscillations are observed in p-n junctions, for  bimetallic ones a plausible mechanism of their generation is proposed in the Appendix. The challenging open question is to apply the similar ideas to systems based on  organic molecules including ``biological engines".\\ 

\textit{Acknowledgments} The support by the Foundation for Polish Science TEAM project co-financed by the EU European Regional Development Fund is acknowledged.

\appendix
\section{Appendix}
In the Appendix the arguments for the existence of slow plasma oscillation modes in bimetallic junctions are presented and the origin of diagonal modulation Hamiltonian is explained for both, semiconductor and bimetallic devices. Here, $\hbar$ is put explicitly in the formulas.
\subsection{Slow charge oscillations in  bimetallic junctions}
In order to find collective oscillating modes of electron gas in a bimetallic junction one can use the one-dimensional hydrodynamical model. The basic variables are: 
the electron density $n(x,t)$, the electrical potential $\psi(x,t)$ and the velocity field $v(x,t)$ satisfying the set of coupled equations
\begin{equation}
m_*\frac{\partial}{\partial t} v + v\frac{\partial}{\partial x} v + \frac{1}{n}\frac{\partial}{\partial x}p - e \frac{\partial}{\partial x}\psi =0
\label{hydro}
\end{equation}
\begin{equation}
\frac{\partial}{\partial t} n + \frac{\partial}{\partial x}(n v) =0
\label{hydro1}
\end{equation}
\begin{equation}
\frac{\partial^2}{\partial x^2} \psi -\frac{e}{\epsilon_0}(n - n_0 ) = 0.
\label{hydro2}
\end{equation}
Here $m_*$ is the effective mass of an electron and $n_0 = n_0(x)$ is the time-independent density of background positive ions (jelium model). The pressure $p(x,t)$ is not an independent variable but is given by the standard formula for the degenerated electron gas
\begin{equation}
p = \frac{(3\pi^2)^{2/3} \hbar^2}{2m_*} n ^{5/3} \equiv \alpha n ^{5/3} .
\label{pressure}
\end{equation}
The stationary solution is given by  $v=0$ and the density $\bar{n}(x)$  which up to a certain smoothing essentially follows the background  density $n_0(x)$.  Consider a small perturbation of the stationary electron gas distribution  in a form of well-localized wave packet, i.e. $ n(x,t) = \bar{n}(x) +\delta n(x,t)$. Inserting also $\psi(x,t) = \bar{\psi}(x) + \delta\psi(x,t)$ into eqs. \eqref{hydro}-\eqref{hydro2} and assuming that the stationary density is a slowly varying function of $x$ in comparison with the variation
of the wave packet $\delta n$ one obtains a set of linearized equations
\begin{equation}
m_*\frac{\partial}{\partial t} v + \frac{5}{3}\frac{\alpha}{\bar{n}^{1/3}}\frac{\partial}{\partial x}\delta n - e \frac{\partial}{\partial x}\delta\psi =0 ,
\label{lhydro}
\end{equation}
\begin{equation}
\frac{\partial}{\partial t} \delta n + \bar{n}\frac{\partial}{\partial x} v = 0 ,
\label{lhydro1}
\end{equation}
\begin{equation}
\frac{\partial^2}{\partial x^2} \delta\psi -\frac{e}{\epsilon_0}\delta n  = 0 .
\label{lhydro2}
\end{equation}

Taking the derivative of eq.\eqref{lhydro} with respect to $x$ then inserting relations \eqref{lhydro1},\eqref{lhydro2} (omitting the derivatives of $\bar{n}$) one obtains 
the following one-dimensional Klein-Gordon-like equation
\begin{equation}
\Bigl[\frac{\partial^2}{\partial t^2}  - {c_F}^2\frac{\partial^2}{\partial x^2} + \omega_p^2 \Bigr]\delta n =0
\label{KG}
\end{equation}
for the density perturbation $\delta n$. Here, the maximal velocity $c_F$ is given by
\begin{equation}
c_F^2 = \frac{5}{3}\frac{(3\pi^2)^{2/3} \hbar^2}{2{m_*}^2} \bar{n}^{2/3} = \frac{5}{6} v_F^2 .
\label{max_v}
\end{equation}
where $v_F$ is the Fermi velocity and $\omega_p$ is given by \eqref{plasma_f}.
\par
For an inhomogeneous material like a bimetallic junction $c_F$ and $\omega_p$ are assumed to be slowly varying functions of $x$. Using the analogy to the relativistic Klein-Gordon equation one concludes that the center of the localized perturbation $\delta n$  moves like a fictitious one-dimensional particle with the ``relativistic" Hamiltonian
\begin{equation}
H(x,p) = \sqrt{c_F^2(x) p^2 + \hbar^2\omega_p^2(x)}.
\label{KGHam}
\end{equation}
The approximate Hamiltonian equations in the ``non-relativistic regime" (small $p$) read
\begin{equation}
\dot{x} = \frac{p}{M},\quad M =\frac{\hbar\omega_p}{c_F^2}
\label{KGHam1}
\end{equation}
\begin{equation}
\dot{p} = -\frac{\partial}{\partial x} U(x) ,\quad  U(x) = \hbar \omega_p(x).
\label{KGHam2}
\end{equation}
To advocate the existence of slow modes in  bimetallic junctions one can propose the  qualitative but plausible shape of the effective potential $U(x)$ which interpolates between two bulk values characteristic for both metals (see Fig. 2). The well in the middle can be explained by the presence of a large number of defects  in the transition region between two different lattices. Those defects trap a certain number of electrons reducing the density of free electrons and hence also the local value of $\omega_p(x)\sim\sqrt{\bar{n}(x)}$. As the jump of the effective potential $\Delta U = \hbar (\omega_p^{(B)} - \omega_p^{(A)})$ is of the order of few $eV$ one can estimate the depth of the well
as $\Delta E = 0.1 - 1 eV$ (still higher than the typical value of $ T_1 \simeq 0.05 eV$) and a reasonable value for its width as $L =100 nm$ (comparable to a typical roughness of  well polished metallic surfaces). Using the harmonic well approximation $U(x) \simeq \frac{1}{2}M\Omega^2 x^2$ one can estimate the oscillation frequency  of the plasmonic wave packet confined in the well by
\begin{equation}
\Omega = \sqrt{\frac{2\Delta E}{M L^2}} .
\label{KGfr}
\end{equation}
Taking typical values  $\omega_p \simeq 10^{15} s^{-1}$ and  $v_F \simeq 10^6 \frac{m}{s}$ one obtains  $M \simeq 10^{-30} kg$ and hence, finally $\Omega \simeq 1 THz$ - the value comparable to that for plasma oscillations in semiconductor p-n junctions.
\begin{figure}[tb]
	\centering
\includegraphics[width=\columnwidth,keepaspectratio]{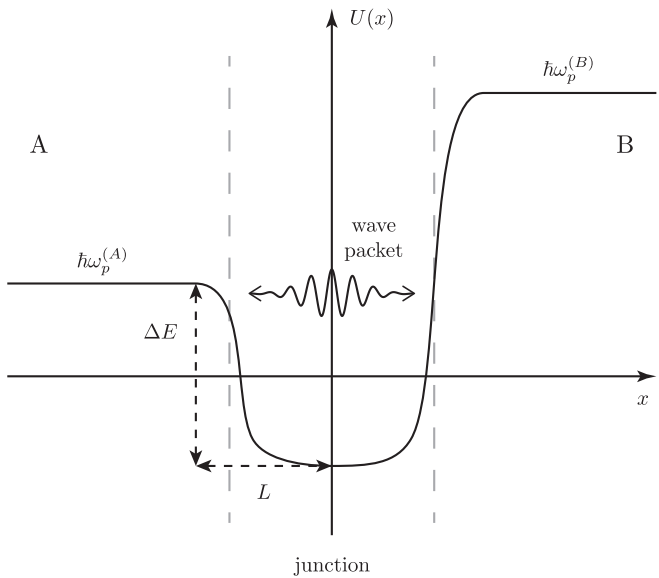}
	\caption{\textbf{Bimetallic junction}. The shape of the effective potential for the plasmonic wave packet. It is instructive to compare the proposed shape of $U(x)$,  proportional to $\sqrt{\bar{n}(x)}$, with the very similar shape of  electron number density obtained numerically in \cite{Ferrante} for the case of Al-Mg contact. In \cite{Ferrante} (FIG.1) the ideal plane surfaces are separated by a small distance of $0.3 nm$. This gap produces physical effects similar to 
the real metal-metal interface.}
	\label{fig:Figure2}
\end{figure}
\subsection{Effective modulation Hamiltonians}
The aim of this section is to justify the choice of the diagonal modulation as a generic consequence of the weak coupling between the system and the driving oscillations.\\

Consider first the case of slow and weak perturbation described by the Hamiltonian $H_0 + \lambda V(t)$ which is applicable to p-n  junction. For a fixed $t$ one can apply  the lowest order of the standard perturbation theory for approximate eigenvectors and eigenenergies
\begin{equation}
\phi_j \simeq \phi^{(0)}_j ,
\label{stan_0}
\end{equation}
\begin{equation}
E_j \simeq E^{(0)}_j + \lambda \langle\phi^{(0)}_j |{V}| \phi^{(0)}_j\rangle.
\label{energia_I}
\end{equation}
Within this approximation one can use the effective time-dependent Hamiltonian with diagonal perturbation
\begin{equation}
H(t) = H_0 + \lambda \sum_{j} \langle\phi^{(0)}_j |{V(t)}| \phi^{(0)}_j \rangle |\phi^{(0)}_j\rangle\langle\phi^{(0)}_j| .
\label{ham_slow}
\end{equation}
For the case of bimetallic junction one should take into account the fact that the source of perturbation  is  a ``wave packet"  with fast ``internal" charge oscillations at the frequency $\omega_p$ and slow oscillatory  motion of its envelope. Hence, the resulting time-dependent Hamiltonian possesses the following structure
\begin{equation}
H(t) = H_0 + \lambda [f(t) \cos\omega_p t]\, V,
\label{ham_fast}
\end{equation}
where $f(t)$ describes the slowly varying envelope. In this case the terms linear in $\lambda$  are practically averaged out. Using the second order of the standard perturbation theory 
\begin{equation}
\phi_j \simeq \phi^{(0)}_j + \lambda \sum_{k\neq j}\frac{\langle \phi^{(0)}_k |{V}|\phi^{(0)}_j \rangle}{E^{(0)}_j - E^{(0)}_k}\, \phi^{(0)}_k.
\label{stan_I}
\end{equation}
\begin{equation}
E_j \simeq E^{(0)}_j + \lambda \langle\phi^{(0)}_j |{V} \phi^{(0)}_j\rangle+ \lambda^2\sum_{k\neq j}\frac{|\langle \phi^{(0)}_k |{V}|\phi^{(0)}_j \rangle|^2}{E^{(0)}_j - E^{(0)}_k},
\label{energia_II}
\end{equation}
and averaging over  fast oscillations one obtains the following effective Hamiltonian
\begin{equation}
H(t) = H_0 + \frac{1}{2}\lambda^2 f^2(t) \sum_{j}\sum_{k\neq j}\frac{|\langle \phi^{(0)}_k |{V}|\phi^{(0)}_j \rangle|^2}{E^{(0)}_j - E^{(0)}_k}|\phi^{(0)}_j\rangle\langle\phi^{(0)}_j| 
\label{ham_fastII}
\end{equation}
with, again, diagonal modulation term.\\

Notice, that for the bimetallic junction the coupling of electrons to charge oscillations is the second order effect and hence, much weaker than for the semiconductor devices. However, this is compensated by  much higher electron densities in metals.

\end{document}